\documentclass[eqsecnum,superscriptaddress,showpacs]{revtex4}
\usepackage{graphics}
\usepackage{latexsym}
\usepackage{bm}
\begin{document}
\title{New Renormalization Scheme of Vacuum Polarization in QED}

\author{Takehisa Fujita}\email{fffujita@phys.cst.nihon-u.ac.jp}
\author{Naohiro Kanda}\email{nkanda@phys.cst.nihon-u.ac.jp}
\author{Hiroshi Kato}\email{hhkato@phys.cst.nihon-u.ac.jp}
\author{Hiroaki Kubo}\email{n-kubo@phys.cst.nihon-u.ac.jp}
\author{Yasunori Munakata}\email{munakata@phys.cst.nihon-u.ac.jp}
\author{Sachiko Oshima}\email{oshima@phys.cst.nihon-u.ac.jp} 
\author{Kazuhiro Tsuda}\email{nobita@phys.cst.nihon-u.ac.jp}
\affiliation{Department of Physics, Faculty of Science and Technology, 
Nihon University, Tokyo, Japan}

\date{\today}%

\begin{abstract}
We examine the vacuum polarization contribution in the renormalization scheme 
of QED. Normally, the quadratic divergence term is discarded under the condition 
that the counter term of the Lagrangian density should be gauge invariant. 
Here, it is shown that the whole contribution of the photon self-energy should not 
be considered for the renormalization procedure. In fact, the finite contribution 
of the renormalization in the vacuum polarization is shown to give rise to 
the hyperfine splitting energy which disagrees with the experimental 
observation in hydrogen atom. For the treatment of the vacuum polarization, we present 
a new renormalization scheme of the photon self-energy diagram.

\end{abstract}


\maketitle

%\vspace{1cm}
\noindent

%\newpage
\section{Introduction}
In the renormalization procedure of QED, one considers the vacuum polarization 
which is the contribution of the self-energy diagram of photon 
$$ \Pi^{\mu \nu}(k)=ie^2\int {d^4p\over(2\pi)^4}
{\rm Tr} \left[ \gamma^\mu {1\over p \llap/-m  } \gamma^\nu 
{1\over p \llap/-k \llap/-m  }\right] .  \eqno{(1.1)} $$
This integral obviously gives rise to the quadratic divergence ($\Lambda^2$ term). 
However, when one considers the counter term of the Lagrangian density which should cancel 
this quadratic divergence term, then the counter Lagrangian density violates the gauge 
invariance since it should correspond to the mass term in the gauge field Lagrangian 
density. Therefore, one has to normally erase it by hand, and in the cutoff procedure of 
the renormalization scheme, one subtracts the quadratic divergence term such that one can 
keep the gauge invariance of the Lagrangian density. Here, we should notice that 
the largest part of the vacuum polarization contributions is discarded, and this indicates 
that there must be something which is not fully understandable in the renormalization 
procedure. Physically, it should be acceptable to throw away the $\Lambda^2$ term since 
this infinite term should not be connected to any physical observables. Nevertheless 
we should think it over why the unphysical infinity appears in the self-energy diagram 
of photon.   

On the other hand, the quadratic divergence term disappears in the treatment of 
the dimensional regularization scheme. Here, we clarify why the quadratic divergence term 
does not appear in the dimensional regularization treatment.  That is, the treatment 
of the dimensional regularization employs the mathematical formula which is not valid 
for the evaluation of the momentum integral. Therefore, the fact that there is no 
quadratic divergence term in the dimensional regularization is simply because one makes 
a mistake by applying the invalid mathematical formula to the momentum integral. This is 
somewhat surprising, but now one sees that the quadratic divergence is still there 
in the dimensional regularization, and this strongly indicates that we should reexamine 
the effect of the photon self-energy diagram itself. 

This paper is organized as follows. In the next section, we discuss the momentum integral 
in the vacuum polarization contributions, and the integration rotated into 
the Euclidean space is presented. In section 3, we treat the dimensional regularization 
and discuss the problem in the application of the gamma function which appears 
in the momentum integral. In section 4, we show the calculated result of 
the hyperfine splitting energy in hydrogen atom which is corrected from the finite 
contribution of the vacuum polarization. In section 5, we discuss the physical 
reason why one should not consider the vacuum polarization diagram 
in the renormalization procedure in QED. Finally, section 6 summarizes what we clarify 
in this paper.

%\vspace{1cm}
\section{Momentum Integral with Cutoff $\Lambda$}
In this section, we briefly review the standard renormalization scheme of 
the vacuum polarization diagram.  
The evaluation of the photon self-energy diagram is well explained in the textbook of 
Bjorken and Drell \cite{bd}, and therefore we describe here the simplest way of calculating 
the momentum integral. The type of integral one has to calculate can be summarized as
$$ \int d^4 p {1\over (p^2-s+i\varepsilon)^n} = {i\pi^2}\int_0^{\Lambda^2} 
w dw {1\over (w-s+i\varepsilon)^n} 
 \ \ \ {\rm with} \ \ \ w=p^2 \eqno{(2.1)}  $$
where $i$ appears because the integral is rotated into the Euclidean space and this 
corresponds to $D=4$ in the dimensional regularization as we will see it below. 

\subsection{Photon Self-energy Contribution $ \Pi^{\mu \nu}(k)$}
The photon self-energy contribution $ \Pi^{\mu \nu}(k)$ 
can be easily evaluated as
$$ \Pi_{\mu \nu}(k)=  {4ie^2\over (2\pi)^4}\int_0^1 dz \int d^4p 
 \left[ {2p_\mu p_\nu- g_{\mu \nu} p^2+sg_{\mu \nu}-2z(2-z)
(k_\mu k_\nu-  k^2g_{\mu \nu})\over{(p^2-s+i\varepsilon)^2}}  \right] $$
$$ ={\alpha\over 2\pi}\int_0^1 dz \int_0^{\Lambda^2} dw  
\left[ { w(w -2s)g_{\mu \nu}+4z(1-z)w
(k_\mu k_\nu-  k^2g_{\mu \nu})\over{(w-s+i\varepsilon)^2}}  \right]  \eqno{(2.2)}  $$
where $s$ is defined as $s=m^2-z(1-z)k^2$. This can be calculated to be
$$ \Pi_{\mu \nu}(k)=\Pi^{(1)}_{\mu \nu}(k)+\Pi^{(2)}_{\mu \nu}(k) $$
where
$$ \hspace{-8cm} \Pi^{(1)}_{\mu \nu}(k)={\alpha \over 2\pi} \left(\Lambda^2+m^2
-{k^2\over 6} \right)g_{\mu \nu}  \eqno{(2.3a)}$$
$$ \Pi^{(2)}_{\mu \nu}(k)={\alpha \over 3\pi}(k_\mu k_\nu- k^2g_{\mu \nu})
\left[ \ln \left({\Lambda^2\over m^2e}\right) -6\int_0^1 dz z(1-z) 
\ln \left(1-{k^2\over m^2}z(1-z) \right) \right] . \eqno{(2.3b)}  $$
Here, the $ \Pi^{(1)}_{\mu \nu}(k)$ term corresponds to the quadratic divergence term 
and this should be discarded since it violates the gauge invariance when one considers 
the counter term of the Lagrangian density. 
The $ \Pi^{(2)}_{\mu \nu}(k)$ term can keep the gauge invariance, and therefore 
one can renormalize it into the new Lagrangian density. 

\subsection{Finite Term in Photon Self-energy Diagram}
After the renormalization, one finds a finite term which should affect the propagator 
change in the process involving the exchange of the transverse photon $\bm{A}$. 
The propagator ${1\over q^2}$ should be replaced by 
$$ {1\over q^2} \Rightarrow {1\over q^2}\left[ 1+{2\alpha \over \pi} 
\int_0^1 dz z(1-z)\ln \left( 1-{q^2z(1-z)\over m^2}\right) \right] \eqno{(2.4)} $$
where $q^2$ should become $q^2 \approx -\bm{q}^2$ for small $\bm{q}^2$. 
It should be important to note that the correction term arising from the finite 
contribution of the photon self-energy diagram should affect only for the renormalization 
of the vector field  $\bm{A}$. Since the Coulomb propagator is not affected by 
the renormalization procedure of the transverse photon (vector field  $\bm{A}$), 
one should not calculate its effect on the Lamb shift where the propagator is, of course, 
the static Coulomb propagator.

%\vspace{1cm}
\section{Dimensional Regularization}
Before discussing the physical effect of the renormalization of the photon self-energy 
contribution, we examine the dimensional regularization method \cite{thv1,thv2} 
which is commonly used in evaluating the momentum integral in self-energy diagrams. 
In the evaluation of the momentum integral which appears in the photon self-energy diagram, 
one often employs the dimensional regularization where the integral is replaced as
$$ \int{d^4p\over (2\pi)^4} \rightarrow \lambda^{4-D}\int{d^Dp\over (2\pi)^D} 
\eqno{(3.1)} $$
where $\lambda$ is introduced as a parameter which has a mass dimension in order to 
compensate the unbalance of the momentum integral dimension. This is the integral 
in the Euclidean space, but $D$ is taken to be 
$D=4-\epsilon$ where $\epsilon$ is an infinitesimally small number. 

\subsection{Photon Self-energy Diagram with $D=4-\epsilon$}
In this case, the photon self-energy $ \Pi_{\mu \nu}(k)$ can be calculated to be
$$ \Pi_{\mu \nu}(k)=i\lambda^{4-D}e^2\int {d^Dp\over(2\pi)^D} 
Tr \left[ \gamma_\mu {1\over p \llap/-m  } \gamma_\nu 
{1\over p \llap/-k \llap/-m  }\right] $$
$$ ={\alpha\over 3\pi}(k_\mu k_\nu-g_{\mu \nu}k^2) \left[ {2\over \epsilon}
+ \textrm{finite term}  \right] \eqno{(3.2)}  $$
where the finite term is just the same as eq.(2.3). In eq.(3.2), one sees that 
the quadratic divergence term ($\Pi^{(1)}_{\mu \nu}(k)$) is missing. This is surprising 
since the quadratic divergence term is the leading order contribution in the momentum 
integral, and whatever one invents in the integral, there is no way to erase it 
unless one makes a mistake. 

\subsection{Mathematical Formula of Integral}
Indeed, in the treatment of the dimensional regularization, 
people employ the mathematical formula which is invalid for the integral in eq.(3.2). 
That is, the integral formula for $D=4-\epsilon$ 
$$ \int d^Dp {p_\mu p_\nu\over{ (p^2-s+i\varepsilon)^n }}
=i\pi^{D\over 2}(-1)^{n+1} {\Gamma(n-{1\over 2}D-1)
\over{2\Gamma(n)}} {g_{\mu \nu}\over s^{n-{1\over 2}D-1} } 
\ \ \ ({\rm for} \ \ \ n \geq 3)  \eqno{(3.3)}  $$
is only valid for $n \geq 3$ in eq.(3.3). For $n=3$, the integral should have 
the logarithmic divergence, and this is nicely avoided by the replacement of 
$D=4-\epsilon$. However, the $n=2$ case must have the quadratic divergence and 
the mathematical formula of eq.(3.3) is meaningless. 
In fact, one should recover the result of the photon self-energy contribution 
$ \Pi_{\mu \nu}(k)$ of eqs.(2.3) at the limit of $D=4$, apart from 
the $(2/ \epsilon)$ term which 
corresponds to the logarithmic divergence term. 

\subsection{Reconsideration of Photon Self-energy Diagram}
In mathematics, one may define the gamma function in terms of the algebraic equations with 
complex variables. However, the integral in the renormalization procedure is defined only 
in real space integral, and the infinity of the integral is originated  from the infinite 
degrees of freedom in the free Fock space \cite{fujita}. 

Therefore, one sees that the disappearance of the quadratic divergence 
term in the evaluation of $\Pi_{\mu \nu}(k)$ in the dimensional 
regularization is not due to the mathematical trick, but simply due to 
a simple-minded mistake. In this respect, it is just accidental that 
the $\Pi^{(1)}_{\mu \nu}(k)$ term in the dimensional regularization vanishes to zero. 
Indeed, one should obtain the same expression of the $\Pi^{(1)}_{\mu \nu}(k)$ term 
as eq.(2.3a) when one makes $\epsilon \rightarrow 0$ in the calculation of 
the dimensional regularization. 
This strongly suggests that we should reconsider the photon self-energy diagram itself 
in the renormalization procedure.

%\vspace{1cm}
\section{Propagator Correction of Photon Self-energy}
In order to examine whether the inclusion of the photon self-energy contribution 
is necessary for the renormalization procedure or not, 
we should consider the effect of the finite contribution from 
the photon self-energy diagram. As we see, there is a finite contribution 
of the transverse photon propagator to physical observables after the renormalization 
of the photon self-energy. The best application of the propagator correction must be 
the magnetic hyperfine splitting of the ground state ($1s_{1\over 2}$ state) 
in the hydrogen atom since this interaction is originated from the vector field $\bm{A}$ 
which gives rise to the magnetic hyperfine interaction between electrons and nucleus. 

\subsection{Lamb Shift Energy}
In some of the textbooks \cite{bd}, the correction term arising from the finite 
contribution of the photon self-energy diagram is applied to the evaluation of 
the Lamb shift energy in hydrogen atom, and it is believed that the finite correction term 
should give the Lamb shift energy of $-27$ MHz in the $2s_{1\over 2}$ state in hydrogen 
atom. However, this is not a proper application 
since only the renormalization of the vector field  $\bm{A}$ should be considered. 
This is closely connected to the understanding of the field quantization itself. 
One sees that the second quantization of the electromagnetic field should be made 
only for the vector field $\bm{A}$, and this is required 
from the experimental observation that photon is created from the vacuum of 
the electromagnetic field in the atomic transitions. 
Therefore, it is clear that the renormalization becomes necessary only 
for the vector field  $\bm{A}$. 

In fact, the Coulomb propagator is 
not affected by the renormalization procedure of the transverse photon since the $A_0$ 
term is exactly solved from the constraint equation. In this respect, the Lamb shift 
energy of the $2s_{1\over 2}$ state in hydrogen atom should be understood without 
the finite correction term of the propagator from the vacuum polarization. 
However, it may well be difficult to calculate the Lamb shift energy to a high accuracy 
since it involves the bound state wave function of the hydrogen atom. At the accuracy  
of $10^{-3}$ level, one has to carefully consider the reduced mass effect which is 
normally neglected in the Lamb shift calculations \cite{bd}. In addition, if one should 
avoid the problem of the logarithmic divergence in the Lamb shift calculation, 
one has to carry out the fully relativistic calculations. However, in this case, 
one has to face the difficulty of the negative energy states in hydrogen atom, and 
this is a non-trivial task.

\subsection{Magnetic Hyperfine Interaction}
The magnetic hyperfine interaction between electron and proton in hydrogen atom 
can be written with the static approximation in the classical field theory as
$$ H'=-\int \bm{j}_e (\bm{r}) \cdot \bm{A} (\bm{r})d^3 r \eqno{(4.1)}  $$
where $\bm{j}_e (\bm{r}) $ denotes the current density of electron, and 
$\bm{A} (\bm{r})$ is the vector potential generated by proton and is given as
$$ \bm{A} (\bm{r})={1\over 4\pi}\int {\bm{J}_p (\bm{r}')\over{|\bm{r}-\bm{r}'|}}d^3r' 
 \eqno{(4.2)}  $$ 
where $\bm{J}_p (\bm{r})$ denotes the current density of proton. 
The hyperfine splitting of the ground state in the hydrogen atom can be calculated as
$$  \Delta E_{hfs}=\langle 1s_{1\over 2},I:F|H'|1s_{1\over 2},I:F \rangle \eqno{(4.3)}  $$
where $I$ and $F$ denote the spins of proton and atomic system, respectively. 
This can be explicitly calculated as
$$  \Delta E_{hfs}=\left(2F(F+1)-3\right) {\alpha g_p\over 3M_p}
\int_0^\infty F^{(1s)}(r)G^{(1s)}(r) dr  \eqno{(4.4)}  $$
where $g_p$ and $M_p$ denote the g-factor and the mass of proton, respectively. 
$F^{(1s)}(r)$ and $G^{(1s)}(r)$ are the small and large components of the radial 
parts of the Dirac wave function of electron in the atom. In the nonrelativistic 
approximation, the integral can be expressed as
$$ \int_0^\infty F^{(1s)}(r)G^{(1s)}(r) dr \simeq {(m_r\alpha)^3\over  m_e } 
 \eqno{(4.5a)}  $$
where $m_e$ is the mass of electron and $m_r$ denotes the reduced mass defined as
$$ m_r={m_e\over{1+{m_e\over M_p}} }.  \eqno{(4.6)}  $$
It should be noted that, in eq.(4.5), $m_e$ appears in the denominator because it is 
originated from the current density of electron. Therefore, 
the energy splitting between $F=1$ and $F=0$ atomic states in the nonrelativistic 
limit with a point nucleus can be calculated from Eq.(4.4) as 
$$ \Delta E_{hfs}^{(0)}=   {8\alpha^4m_r^3\over 3m_e M_p}.   \eqno{(4.7)} $$

\subsection{QED Corrections for Hyperfine Splitting}
There are several corrections which arise from the various QED effects such as 
the anomalous magnetic moments of electron and proton, nuclear recoil effects and 
relativistic effects. We write the result 
$$ \Delta E_{hfs}^{(QED)}=   {4g_p\alpha^4m_r^3\over 3m_e M_p}   
(1+a_e)\left(1+{3\over 2}\alpha^2\right) (1+\delta_R)  \eqno{(4.8)} $$
where $a_e$ denotes the anomalous magnetic moment of electron. 
The term $\left(1+{3\over 2}\alpha^2\right)$ 
appears because of the relativistic correction of the electron wave function
$$ \int_0^\infty F^{(1s)}(r)G^{(1s)}(r) dr = {(m_r\alpha)^3\over  m_e } 
\left(1+{3\over 2}\alpha^2 + \cdots \right) . \eqno{(4.5b)}  $$
The term $\delta_R$ corresponds to the recoil corrections and can be written as 
\cite{bodwin}
$$ \delta_R =\alpha^2 \left(\ln 2-{5\over 2}\right)-{8\alpha^3\over 3\pi}
\ln \alpha \left(\ln \alpha-\ln 4+{281\over 480}\right)+
{15.4\alpha^3\over \pi} .  \eqno{(4.9)}  $$  
Now the observed value of $\Delta E_{hfs}^{(exp)}$ is found to be \cite{hfsexp}
$$   \Delta E_{hfs}^{(exp)}= 1420.405751767 \  {\rm MHz} . $$
Also, we can calculate $\Delta E_{hfs}^{(0)} $ and $\Delta E_{hfs}^{(QED)} $ 
numerically and their values become
$$ \Delta E_{hfs}^{(0)}  = 1418.83712 \ \  {\rm MHz}, \ \ \ \ 
\Delta E_{hfs}^{(QED)} = 1420.448815 \ \  {\rm MHz} . $$
Therefore, we find the deviation from the experimental value as
$$ {\Delta E_{hfs}^{(exp)}-\Delta E_{hfs}^{(QED)}
\over \Delta E_{hfs}^{(0)} } \simeq -30 \ \ {\rm ppm} .  \eqno{(4.10)}  $$

\subsection{Finite Size Corrections for Hyperfine Splitting}
In addition to the QED corrections, there is a finite size correction of proton and 
its effect can be written as
$$ \Delta E_{hfs}^{(FS)}= \Delta E_{hfs}^{(0)}(1 +\varepsilon)  \eqno{(4.11)} $$  
where the $\varepsilon$ term  corresponds to the Bohr-Weisskopf effect 
\cite{bohr,zemach,tfhfs}
$$ \varepsilon \simeq -m_e \alpha R_p  \eqno{(4.12)} $$
where $R_p$ denotes the radius of proton. It should be noted that the perturbative 
treatment of the finite proton size effect on the hyperfine splitting overestimates 
the correction by a factor of two. Now, the calculated value of 
$\varepsilon$ becomes 
$$\varepsilon \simeq -17 \ \ {\rm ppm}. $$ 
Therefore, the agreement between theory and experiment is quite good.

\subsection{Finite Propagator Correction from Photon Self-energy}
The hyperfine splitting of the $1s_{1\over 2}$ state energy 
including the propagator correction can be written in terms of the momentum 
representation in the nonrelativistic limit as 
$$ \Delta E_{hfs}^{(VP)}= \left(2F(F+1)-3\right) 
{16\over 3\pi}{\alpha^5m_r^4  \over m_e M_p} 
\int_0^\infty {q^2dq \over{\left(q^2+4(m_r\alpha)^2 \right)^2 }}
\left(1+ M^R(q)\right) $$
$$ \hspace{-6.5cm} \equiv \Delta E_{hfs}^{(0)} (1+\delta_{vp} ). \eqno{(4.13)}  $$
$M^R(q)$ denotes the propagator correction and can be written as
$$ M^R(q)={2\alpha \over \pi} \int_0^1 dz z(1-z) \ln \left( 1+ 
{{q}^2\over m_e^2}z(1-z) \right) .  \eqno{(4.14)}  $$
We can carry out numerical calculations of the finite term of 
the renormalization in the photon self-energy diagram, and we find 
$$ \delta_{vp} \simeq 18\ {\rm ppm} \eqno{(4.15)} $$
which tends to make a deviation larger between theory and experiment of hyperfine 
splitting in hydrogen atom. This suggests that the finite correction from 
the photon self-energy contribution should not be considered for the renormalization procedure.

%\vspace{1cm}
\section{New Renormalization Scheme of Photon Self-energy in QED}
The difficulty of the photon self-energy must be connected to the field quantization 
itself since we do not fully understand it yet. It is clear that the field quantization 
is required from experiment, and the field quantization procedure itself must be 
well justified. However, this does not mean that we understand it theoretically 
within the framework of quantum field theory. For the quantization of 
the vector field $\bm{A} (\bm{r},t)$, one has to first fix the gauge since otherwise one 
cannot determine the gauge field $ \bm{A} (\bm{r},t)$ which depends on the choice 
of the gauge. Then, one can quantize it and accordingly one can quantize 
the Hamiltonian of the electromagnetic field. In this case, one can calculate 
the contributions of the photon self-energy diagram which, in fact, give rise 
to the quadratic divergence. Since this divergence is too difficult to handle, 
people assumed that it should be discarded since it is indeed the same term as 
the mass term of the electromagnetic field Hamiltonian. Therefore, it is required 
that this quadratic divergence term should be discarded by the condition 
that the renormalized Hamiltonian should be gauge invariant. However, one can 
easily notice that this requirement is somewhat incomprehensible since one has 
already fixed the gauge before the field quantization. 

\subsection{Intuitive Picture of No Renormalization of Photon Self-energy}
Now, we should start from the equations of motion for the gauge field as well as 
the fermion field
$$ \Box \bm{A}(x)  = e\bar{\psi}(x)\bm{\gamma}\psi(x)    \eqno{(5.1a)}  $$
$$ \left(i\partial_\mu \gamma^\mu-m \right)\psi(x)
=  e{A}_\mu(x)\gamma^\mu\psi(x) 
  \eqno{(5.1b)}  $$
where  $x$ denotes $x=(t,\bm{r})$. 
These equations of motion can be solved perturbatively by assuming that 
the coupling constant $e$ is small, which is, indeed, an observed fact.  
Here, one sees that the vector field $\bm{A} (x)$ should 
not be influenced by the renormalization procedure since there is no term 
which involves the vector field $\bm{A} (x)$ 
in the right hand side of eq.(5.1a). On the other hand, the fermion field 
$\psi(x)$ should be affected by the perturbative evaluation since the interaction 
term contains the fermion field $\psi(x)$ in the right hand side. Therefore, 
the contribution of the fermion self-energy diagram should be renormalized 
into the fermion mass and the wave function $\psi(x)$. 

%\newpage
\subsection{Integral Equations}
To see it more explicitly, we can convert eqs.(5.1) into the following 
integral equations 
$$ \bm{A}(x) =\bm{A}_0(x) - e\int 
{ e^{iqx}\over q^2 }\left(\widetilde{\bar{\psi}}(q)\bm{\gamma}\widetilde{\psi}(q) 
\right){d^4q\over{(2\pi)^4}}    \eqno{(5.2a)}  $$
$$ \psi(x)=\psi_0(x) - e \int { e^{iqx}\over q \llap/ -m } 
\left(\widetilde{\bm{A}}(q)\cdot\bm{\gamma}\right) \widetilde{\psi}(q) {d^4q\over{(2\pi)^4}} 
\eqno{(5.2b)}  $$
where $\widetilde{\psi}(q)$ and $\widetilde{\bm{A}}(q)$ denote the fields 
in momentum representation. 
Here, $ \bm{A}_0(x)$ and $\psi_0(x)$ denote the free state solutions which satisfy 
$$ \Box \bm{A}_0(x) =0, \ \ \ \ \  
\left(i\partial_\mu \gamma^\mu-m \right)\psi_0(x)=0 . \eqno{(5.3)}  $$

\subsection{S-matrix Evaluation}
If one carries out the calculation of the photon self-energy diagram in the S-matrix 
method, then one finds that the contribution has the quadratic divergence. 
However, the process is not physical one, and therefore one should just keep them 
as the mathematical effects. This can sometimes happen to the S-matrix evaluation 
since the contributions of some of the diagrams include the processes which cannot 
be realized as the physical process. 
In fact, the photon self-energy diagrams should be unphysical since the photon after 
the interaction does not change its quantum state at all. This situation may become 
physically detectable if the photon could be found in the bound state. However, as one 
knows, there is no possibility that photon can be bound. Therefore, there is no chance 
to observe any effects of the photon self-energy contributions, contrary to the fermion 
self-energy contributions where fermions can be found as a bound state which is completely 
different from the free Fock space evaluations.  

\section{Conclusions}
The renormalization procedure in QED is most successful and reliable in field 
theory calculations. In particular, the treatment of the fermion self-energy is 
well established, and the finite effect after the renormalization procedure 
is successfully compared to the experimental observation of the Lamb shift energy of 
the $2s_{1\over 2}$ state in hydrogen atom. Further, the vertex correction 
is well calculated to a high accuracy and is compared to the experimental 
observation of the magnetic moment of electron ($g-2$ experiment), and the agreement 
between theory and experiment is surprisingly good. 
For the above two contributions which have the logarithmic divergences, there is 
no conceptual problem since both of the diagrams are related to the physical processes. 
Further, the physical meaning of the renormalization is well justified even though 
there appears some infinity. This is connected to the fact that the logarithmic 
divergence can never get to a real infinity in physical space, and therefore 
the procedure of the renormalization of the logarithmic infinity has no intrinsic 
difficulty. 

On the other hand, the vacuum polarization contribution is completely 
different from the above two effects. It has the quadratic divergence which is 
impossible to handle for the renormalization procedure. Therefore, before 
the renormalization of the vacuum polarization diagram, one has to discard the photon 
self-energy contribution at the level of the calculation of $ \Pi^{\mu \nu}(k)$. 
This treatment itself cannot be justified within the renormalization scheme of QED. 
Therefore, people had to stick to the gauge invariance of the calculated result 
in order to find any excuse of discarding the quadratic divergence term even though 
the calculated result is obtained by having fixed the gauge.  

In this sense, there must have been many physicists who had an uneasy feeling 
on the treatment of the photon self-energy diagram. The principle we should take 
in physics is that any theoretical frameworks must be connected to physical observables. 
The self-energy of fermion cannot be related to physical observables as far as 
one works within the free Fock space of QED. However, as one knows, fermions can 
become a bound state which is not found in the free Fock space, and therefore 
the physical effect of the self-energy of fermion after the renormalization 
procedure can be detected as the Lamb shift energy in hydrogen atom. 
On the other hand, there is no detectable effect of the self-energy of photon 
since photon can be found always as a free state which is indeed within the free Fock 
space. Therefore, from the beginning, physically there is no need of renormalization 
of self-energy of photon even though mathematically the S-matrix evaluation gives rise 
to the infinite contributions from the self-energy of photon diagram.

\vspace{2cm}
%\newpage

\end{document}